\documentclass[preprint,5p,twocolumn]{elsarticle}

\usepackage[utf8]{inputenc}
\usepackage{stfloats}
\usepackage{lipsum}
\usepackage{amsmath}

\journal{Astroparticle Physics}

\def\CAMOO{${}^{40}$Ca$^{100}$MoO$_{4}$}

\def\sm{$\sim$}

\def\UNIT{ckky}
\def\znbb{$0\nu\beta\beta$}

\newcommand\geniso[2]{\ensuremath{\rm ^{#2}#1}}
\newcommand\newiso[3][]{
\ifx&#1&
  \expandafter\newcommand\csname #2\endcsname[1][#3]{\geniso{#2}{##1}}
\else
  \expandafter\newcommand\csname #1\endcsname[1][#3]{\geniso{#2}{##1}}
\fi
}
\newiso{K}{40}
\newiso{Th}{232}
\newiso{U}{238}
\newiso{Rn}{222}
\newiso{Ra}{226}
\newiso{Pb}{208}
\newiso{Tl}{208}
\newiso{Bi}{214}
\newiso{Ac}{226}
\newiso{Po}{210}
\newiso{Mo}{100}

\begin{document}

\begin{frontmatter}



\title{Neutron and muon-induced background studies for the AMoRE double-beta decay experiment}

\author[addr1]{H.W.~Bae}
\author[addr2]{E.J.~Jeon\corref{cor1}}
\ead{ejjeon@ibs.re.kr}
\author[addr2]{Y.D.~Kim}
\author[addr1]{S.W.~Lee}

\address[addr1]{Department of Physics, Kyungpook National University, Daegu 41566, Republic of Korea}	
\address[addr2]{Center for Underground Physics, Institute for Basic Science (IBS), Daejeon 34126, Republic of Korea}

\cortext[cor1]{Corresponding author}

\begin{abstract}
AMoRE (Advanced Mo-based Rare process Experiment) is an experiment to search a neutrinoless double-beta decay of \Mo~in molybdate crystals. The neutron and muon-induced backgrounds are crucial to obtain the zero-background level ($\textless10^{-5}$~counts/(keV$\cdot$kg$\cdot$yr)) for the AMoRE-II experiment, which is the second phase of the AMoRE project, planned to run at YEMI underground laboratory. To evaluate the effects of neutron and muon-induced backgrounds, we performed Geant4 Monte Carlo simulations and studied a shielding strategy for the AMORE-II experiment. Neutron-induced backgrounds were also included in the study. In this paper, we estimated the background level in the presence of possible shielding structures, which meet the background requirement for the AMoRE-II experiment.  
\end{abstract}


\end{frontmatter}


\section{Introduction}
\label{intro}
AMoRE (Advanced Mo-based Rare process Experiment) \cite{Bhang:2012gn} is an experiment to search for neutrinoless double-beta decays~(\znbb)~ of \Mo~nuclei by using scintillating molybdate crystals, operating at milli-Kelvin temperatures. The second phase of the AMoRE project (AMoRE-II) is being planned to operate at the YEMI underground laboratory (YEMI), located at Handuk, an active iron mine in the region of Mt. Yemi, South Korea~\cite{Alenkov:2015dic}. 
The AMoRE-II experiment will use a \sm200 kg array of molybdate crystals 
with the aim of achieving the zero-background level of $\textless10^{-5}$~counts/(keV$\cdot$kg$\cdot$yr) (\UNIT) in the region of interest (ROI), 3.034$\pm$0.010~MeV. The ROI is given by the Q value of \znbb~of $^{100}$Mo in molybdate crystals, which is at 3034.40(17)~keV~\cite{Rahaman:2008}, and the energy range based on the energy resolution of the detector~\cite{Kim:2015pua}. 
  
Therefore, it is important to precisely understand the effects of background sources on underground experiments of rare events such as \znbb~for the suppression of backgrounds to the aimed level. 
Radiations from radioisotopes in the $^{238}$U,  $^{232}$Th,  $^{40}$K, and  $^{235}$U decay chains in detectors, materials in the nearby detector system, shielding materials, and the rock walls surrounding the experimental enclosure are possible sources of backgrounds. We have studied the impact of backgrounds due to these sources for AMoRE-I that is the first phase of the AMoRE project and found that these backgrounds can be suppressed by the optimized detector design and specific analysis methods, as reported in reference~\cite{Luqman:2017amore}.       

However there are backgrounds due to cosmic rays and neutrons. 
Neutrons are generated by the natural radioactivity of materials in underground environment or induced by cosmic-ray muons, as well as their secondaries. 
Gamma rays are also produced by neutrons and muons by (n,~n$^\prime\gamma$), (n,~$\gamma$), and \textit{bremsstrahlung} processes. The electrons, positrons, and gamma rays produced when muons lose their energy by ionization and radiation, initiate an electromagnetic shower, if they have high enough energy. This shower generates more gamma rays through \textit{bremsstrahlung}. 

In this paper, we quantitatively studied the effects of 
neutron and muon-induced backgrounds on the AMoRE underground experiment by performing Geant4 Monte Carlo simulations. The text is structured as follows: Section~\ref{sec:2} illustrates the detector geometry used for the AMoRE-II simulation. In Sect.~\ref{sec:3}, we simulate the cosmic-muon and muon-induced background to study their effects on the background level, using the Muon Veto system in two different shielding configurations. Neutron-induced backgrounds were also included in the study. Section~\ref{sec:4} describes the simulation results used to quantify the effects of neutron backgrounds, from underground environment, by considering the shielding configurations. Finally, in Sect.~\ref{sec:conclusion}, we conclude that the aimed background level could be achieved with the optimal shielding design for AMoRE-II experiment. 

\begin{figure*}[ht]
\begin{center}
\begin{tabular}{ccc}
\includegraphics[width=0.39\textwidth]{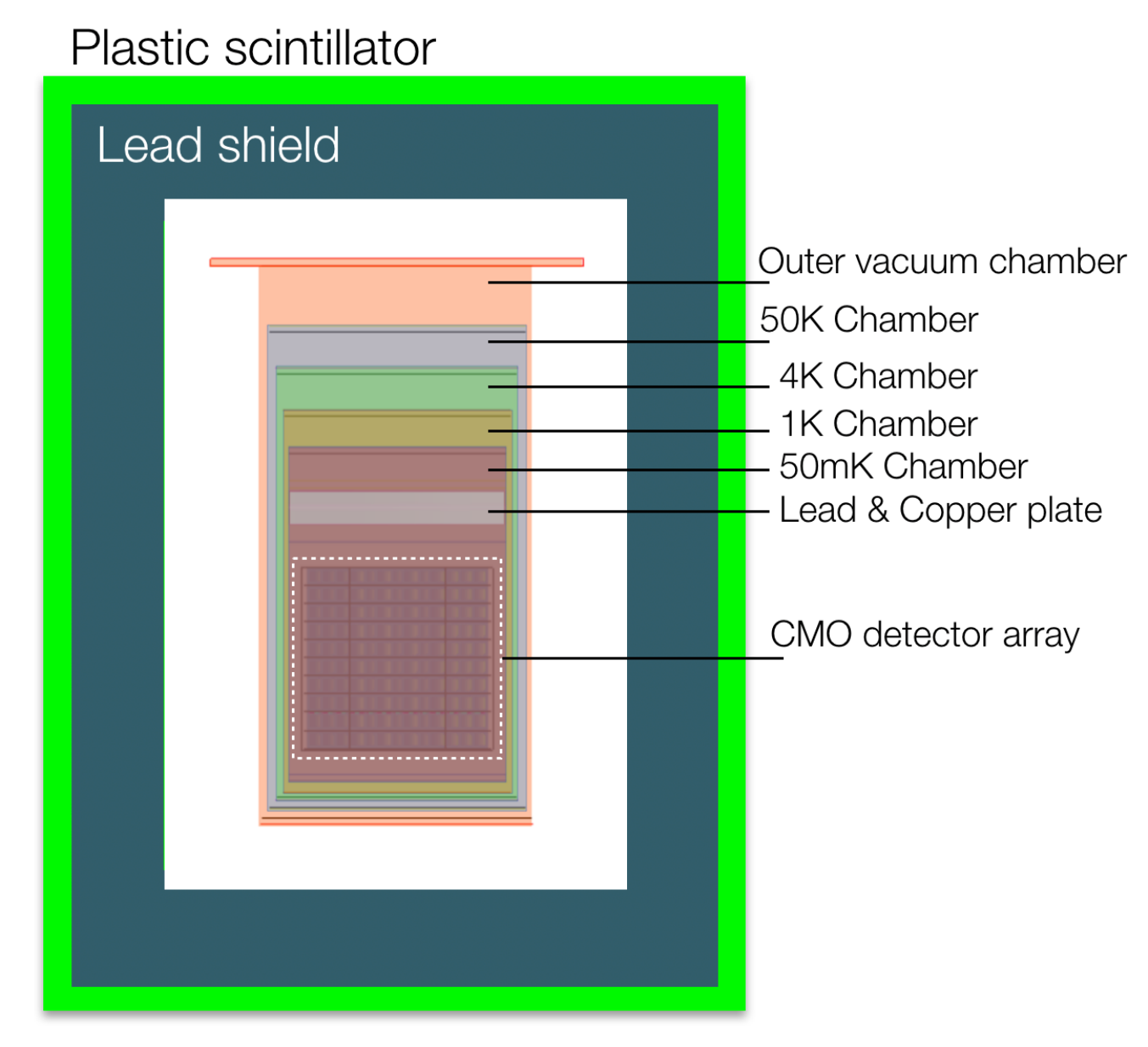} &
\includegraphics[width=0.2\textwidth]{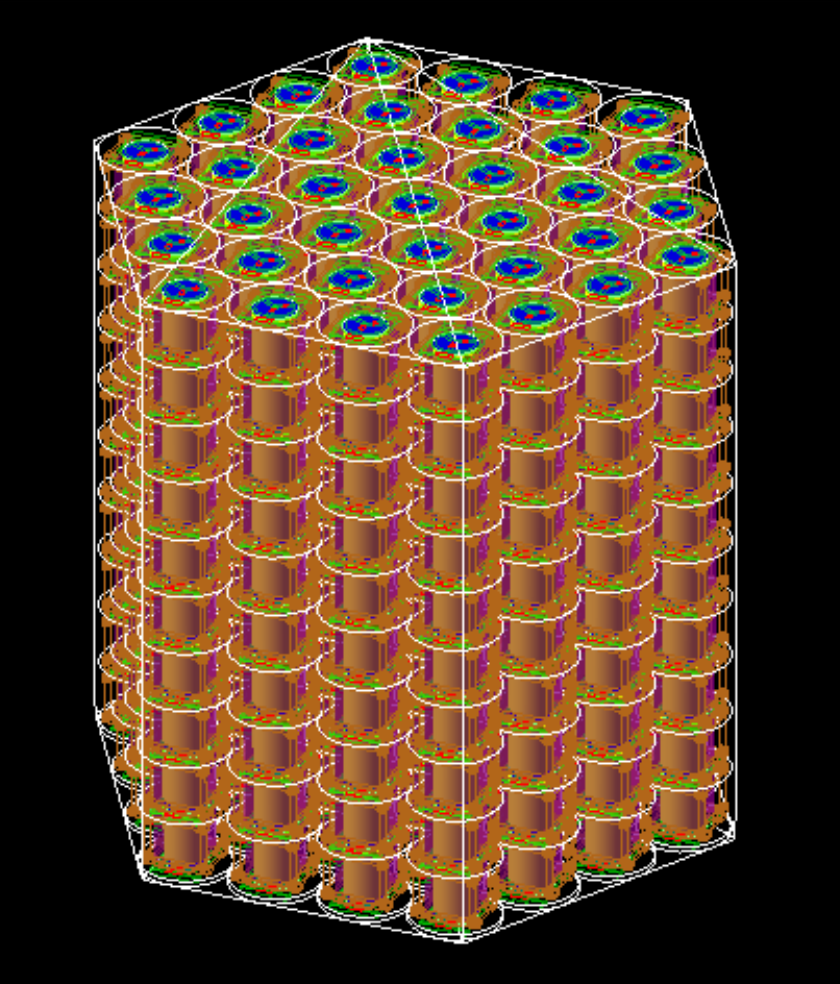} &
\includegraphics[width=0.2\textwidth]{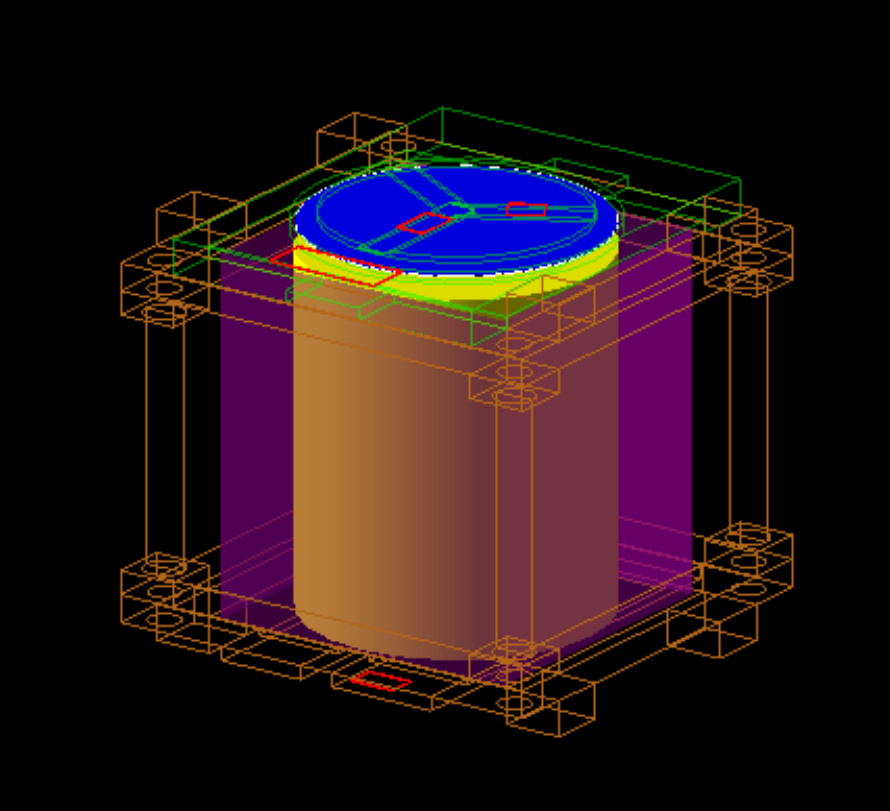} \\
(a) & (b)  & (c) \\
\end{tabular}
\caption{(a) Cross-sectional view of the detector geometry used in the Geant4 simulations. The cryostat is located inside the external lead shield surrounded by 5-cm-thick plastic scintillator~(green). It includes (b) \CAMOO~building and (c) \CAMOO~assembly.
}
\label{detector-setup}
\end{center}
\end{figure*}

\section{Geant4 simulation}
\label{sec:2}
For the background studies, we have performed simulations with the Geant4 Toolkit \cite{Agostinelli:2002hh}. 
We simulated both muons and neutrons using the Geant4 version 4.9.6.p04, in which we implemented the AMoRE-specific physics list. The list was customized for improving the speed and precision of the hadron simulations, over a wide energy range from a few eV to a few hundreds of GeV. It should be highlighted that we also performed the same muon simulations using the higher Geant4 version (10.04.p02) and little difference was found between them. 
We adopted the QGSP$_{-}$BERT$_{-}$HP reference physics list, which is the appropriate one for high-energy physics simulations (above 10~GeV). The precision of the neutron model was also implemented in the physics list for neutrons with energies below 20~MeV. In addition, we used a full-elastic-scattering dataset for thermal neutrons with energies below 4~eV in order to precisely examine the shielding effect.

When a hit occurs, each simulated event included energy depositions in the crystals within a 100-ms event window, which is a few times the typical pulse width ($\sim$20--30 ms) in cryogenic measurements~\cite{Lee:2015tsa}.

\subsection{The AMoRE-II detector geometry}
\label{subsec:2.1}
Fig.~\ref{detector-setup}(a) shows the simplified detector geometry that was used for this study. The cryostat is located inside a 30-cm-thick external lead shield surrounded by 5-cm-thick plastic scintillator. It includes 370 calcium molybdate~(\CAMOO) crystals enclosed by, from the inside outward, a 2-mm-thick magnetic shielding layer of superconducting lead, four copper shielding layers (50mK, 1K, 4K, and 50K chambers) with total thickness of 10 mm, and a 5-mm-thick stainless-steel layer (outer vacuum chamber) of the cryostat. The 370 \CAMOO~(CMO) crystals are arranged in thirty-seven columns, each with ten crystals stacked coaxially, as shown in Fig.~\ref{detector-setup}(b). A 1-cm-thick copper plate and a 10-cm-thick lead plate are located above the CMO crystal array, inside the innermost Cu shield (50mK chamber). Each crystal has a cylindrical shape with 6 cm of diameter, 6 cm of height, and mass of 538 g, resulting in a total mass of 199 kg for all the 370 crystals. Each crystal is covered by a 65~$\mu$m-thick Vikuiti Enhanced Specular Reflector film~\cite{esr} and is mounted in a copper frame. A crystal assembly is shown in Fig.~\ref{detector-setup}(c).

\section{Cosmic muon and muon-induced background}
\label{sec:3}
To quantify the effects of all backgrounds not only from muons but also from the secondary ones induced by muons, traveling through the rocky cavern, the shield, and detector materials, we performed Geant4 Monte Carlo simulations by brute force, starting with muons from the rocks surrounding the cavern. The details of the simulations are described in Sects.~\ref{subsec:3.1} and ~\ref{subsec:3.2}.  

\subsection{Muon energy spectrum at YEMI underground laboratory}
\label{subsec:3.1}
\begin{figure*}[ht]
\begin{center}
\begin{tabular}{cc}
\multicolumn{2}{c}{\includegraphics[width=0.45\textwidth]{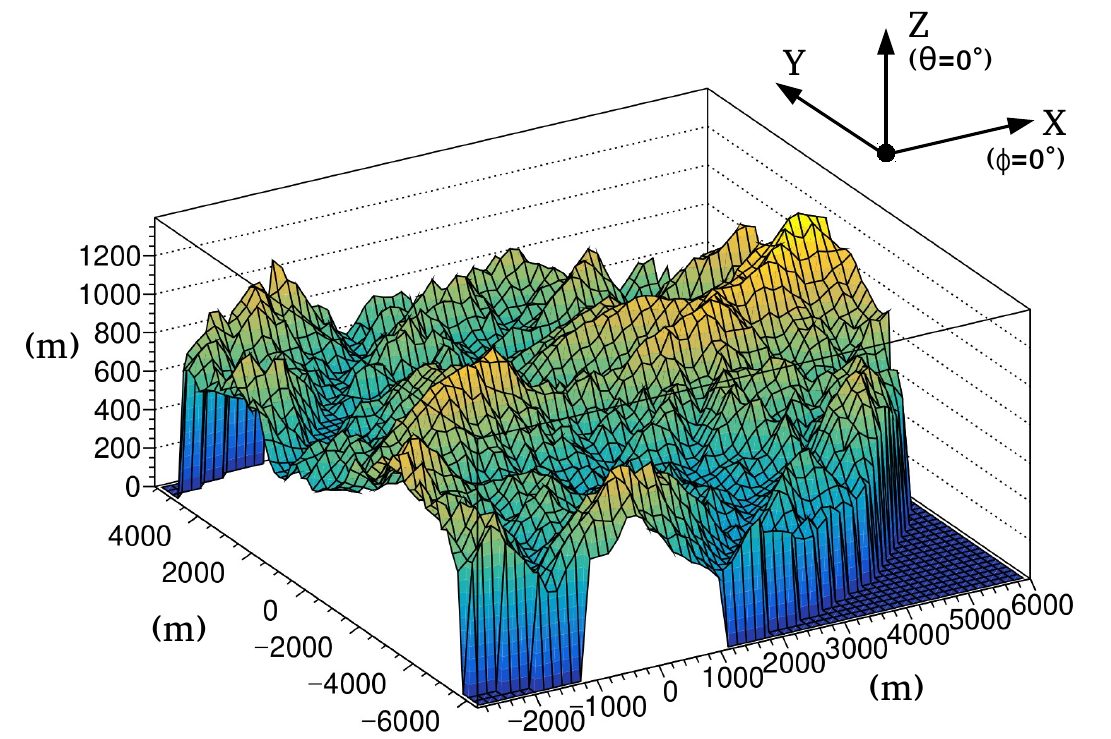}} \\
\multicolumn{2}{c}{(a)} \\
\includegraphics[width=0.45\textwidth]{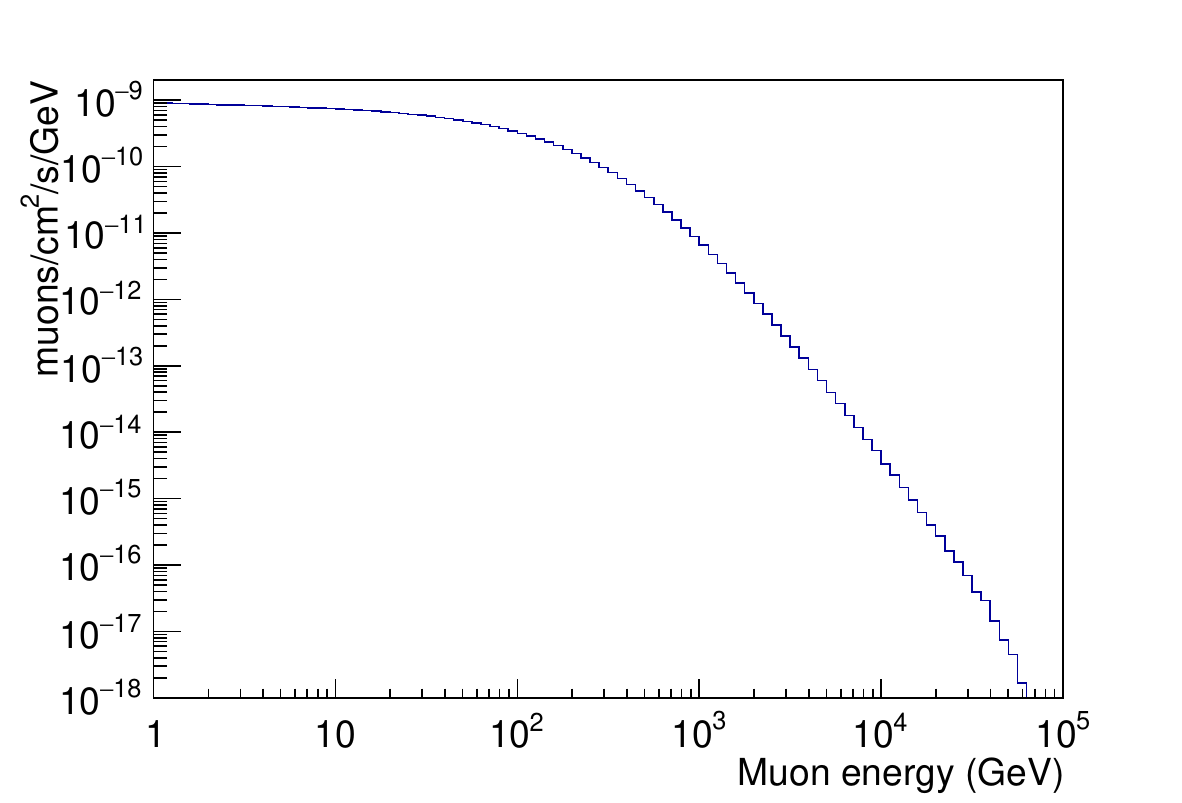} &
\includegraphics[width=0.4\textwidth]{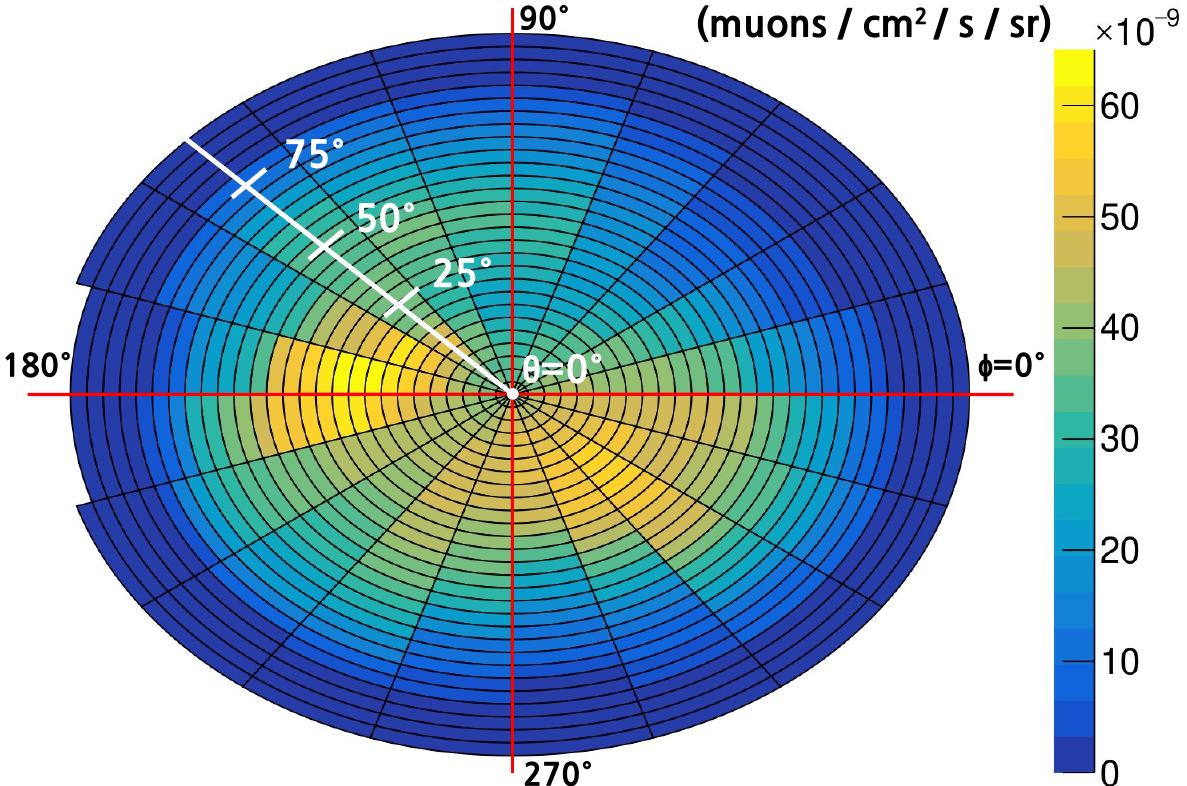} \\
(b)  & (c) \\
\end{tabular}
\caption{(a) Digitized contour map of the YEMI area on a meter scale; the detector is located at (0,0,0), (b) the muon energy spectrum at the YEMI underground laboratory; the mean muon energy is 236~GeV, and (c) the muon intensity in the units of muons/cm$^{2}$/s/sr 
that depends on the azimuthal angle $\phi$ and the polar angle $\theta$ at YEMI.
}
\label{muon-spectrum}
\end{center}
\end{figure*}

The AMoRE-II experiment is the second phase of the AMoRE project planned to run at YEMI underground laboratory (YEMI), located at the Handuk mine, an active iron mine in the region of Mt. Yemi, South Korea. In order to obtain the muon's energy spectrum and the angular distribution at YEMI, we firstly used the sea-level muon flux, parameterized by the modified Gaisser's formula (Section~\ref{subsubsec:3.1.1}). Secondly, we estimated how much energy the muon loses when propagating through the mountain, and the resultant muon energy spectrum to be detected underground, by considering the digitized contour map of the Mt. Yemi area and the parameterized average loss of muon energy within matter (Section~\ref{subsubsec:3.1.2}).

\subsubsection{Modified Gaisser parameterization}
\label{subsubsec:3.1.1}
It is well known that the differential muon intensity at sea level is described by the modified Gaisser parameterization~\cite{modifiedgaisser2006},

\begin{align}
\frac{dN}{dEd\Omega} &= 
	\begin{aligned}[t]
		& A \frac{0.14E^{-2.7}}{\rm cm^2 \cdot s \cdot sr \cdot GeV} \\
		& \times \left(\frac{1}{1+\frac{1.1\tilde{E}\cos\theta^{*}}{115\,{\rm GeV}}}+
		\frac{0.054}{1+\frac{1.1\tilde{E}\cos\theta^{*}}{850\,{\rm GeV}}}+r_c \right), 
	\end{aligned}
\label{eq:modgaisser}
\end{align}

\begin{align}
\cos\theta^{*} &=
	\begin{aligned}[b]	
		& \sqrt{\frac{(\cos\theta)^2+P_1^2+P_2(\cos\theta)^{P_3}
 		+P_4(\cos\theta)^{P_5}}{1+P_1^2+P_2+P_4}}~,
	\end{aligned}	
\label{eq2}
\end{align}
where $P_1$ = 0.102573, $P_2$ = -0.068287, $P_3$ = 0.958633, $P_4$ = 0.0407253, and $P_5$ = 0.817285.

We applied modifications of the parameters A, $r_{c}$, and $\tilde{E}$ for low-energy muons ($E \leq 100/\cos\theta^{*}$), as suggested in Ref.~\cite{modifiedgaisser2006}, which are limited in the standard Gaisser parameterization~\cite{gaisser, gaisser2}. We also took the curvature of the earth into account by using $\cos\theta^{*}$, describing the muon flux for the full range of zenith angles. We generated muon events based on Eq.~(\ref{eq:modgaisser}) with the adequate modified parameters for both high- and low-energy muons. 
   
\subsubsection{YEMI underground laboratory}
\label{subsubsec:3.1.2}
YEMI has a vertical overburdened rock of 1005~meter in the region of Mt. Yemi. A 3-dimensional profile of the mountain area is shown in Fig.~\ref{muon-spectrum}(a), generated by using a digitized contour map of Mt. Yemi area, from Korea Geodetic Datum 2002, based on ITRF2000~\cite{contour}; the (x,y) coordinates of digitized points indicate the location and z-axis is the altitude and the center of the bottom of the detector is located at (0,0,0). Hence, muons generated at sea level by Eq.~(\ref{eq:modgaisser}) lose their energy when propagating through the mountain. We used the formula in Eq.~(\ref{eq:3}) for the average energy loss of a muon traveling a distance $X$ inside the matter, which is parameterized with Eq.~(\ref{eq:4}). The details of the calculation are reported in reference~\cite{muonenergyloss}.

\begin{equation}
- \frac{dE_{\mu}}{dX} = a(E_{\mu}) \,+\, 
\displaystyle\sum_{n=1}^{3} b_{n}(E_{\mu}) \cdot E_{\mu}
\label{eq:3}
\end{equation}
where, 
\begin{eqnarray}
a(E_{\mu}) & = & A_{0} \,+\, (A_{1} \cdot \log_{10}E_{\mu}[GeV]), \nonumber \\ 
\Sigma b(E_{\mu}) & = & B_{0} \,+\, (B_{1} \cdot \log_{10}E_{\mu}[GeV]) \,+\, 
\nonumber \\
 & & \quad \, \{B_{2} \cdot (\log_{10}E_{\mu}[GeV])^2\}
\label{eq:4}
\end{eqnarray}

\begin{figure}[t]
\begin{center}
\includegraphics[width=0.45\textwidth]{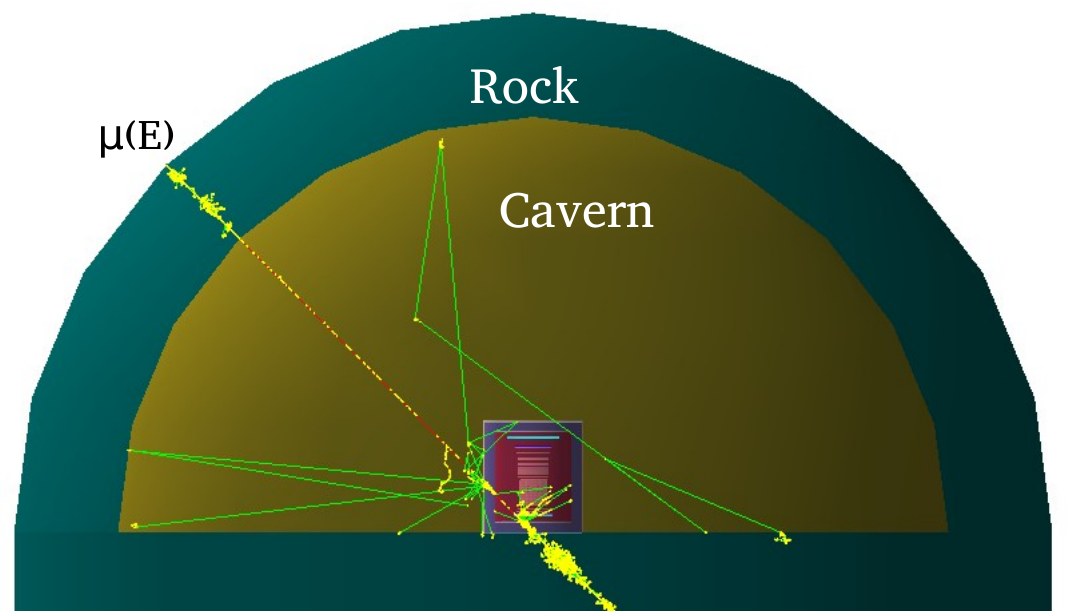}
\caption{
Schematic view of the muon generated by brute-force simulations.
}
\label{bruteforcesim}
\end{center}
\end{figure}
\begin{figure*}[!b]
\begin{center}
\begin{tabular}{cc}
\includegraphics[width=0.5\textwidth]{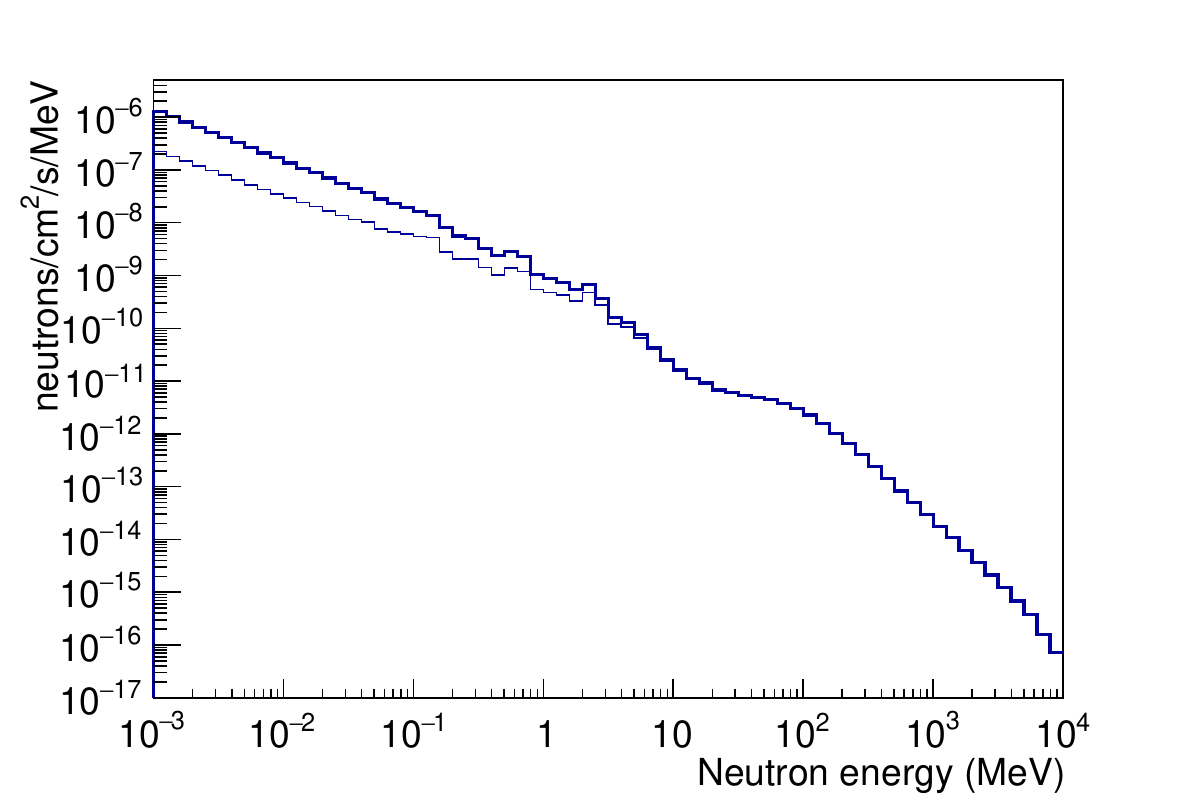} &
\includegraphics[width=0.5\textwidth]{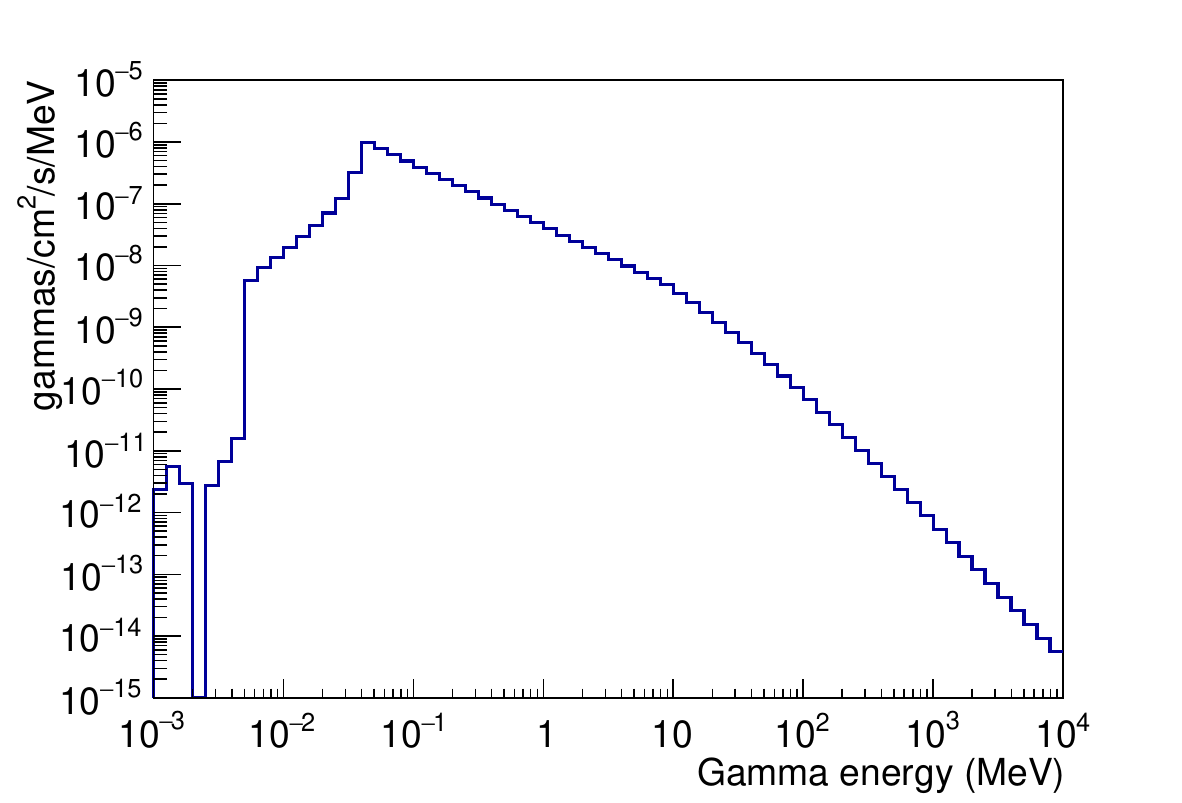} \\
(a) & (b)  \\
\end{tabular}
\caption{
Muon-induced neutrons (a) and gamma rays (b) at the rock surface. The thick solid line of (a) includes the neutrons backscattered from the rock surface, as well as from shielding materials surrounding the detector that reenter the cavern. Neutrons entering the cavern for the first time are represented by the thin line of (a). 
}
\label{energyatrocksurface}
\end{center}
\end{figure*}

We used the following coefficients for the average energy loss in standard rock with the density of 2.65~g/cm$^{3}$: $A_0$~=~1.925 MeV, $A_1$~= ~0.252~cm$^2$/g, and ($B_0$~=~0.358, $B_1$~=~1.711, and $B_2$~=~-0.17), given in the unit of 10$^{-6}$~cm$^2$/g.

We first generated a muon based on the differential muon intensity given by Eq.~(\ref{eq:modgaisser}) 
and then calculated the energy lost by the muon with a path length that starts at the point where the muon track intersects the mountain, expressed by Eq.~(\ref{eq:3}). We assumed that the muon travels in the same direction of its incidence onto the mountain until it reaches the underground laboratory. Accordingly, we obtained the muon energy spectrum to be detected underground and 
the muon intensity in the units of muons/cm$^{2}$/s/sr 
that depends on the azimuthal angle $\phi$ and the polar angle $\theta$ at the YEMI underground laboratory
and the results are shown in Fig.~\ref{muon-spectrum}(b),(c), respectively. 
It is difficult to calculate the absolute muon flux underground due to many ambiguities from rock properties, different depths, etc. Therefore, the integrated muon intensity (through a horizontal area) at YEMI 
can be normalized by the measured flux. 
The total muon flux at YEMI is thus considered as 8.2$\times10^{-8}$~muons/cm$^{2}$/s, which is derived by using the flux of 328$\pm$1(stat)$\pm$10(syst)~muons/m$^2$/day, measured  at the Yangyang underground laboratory~(Y2L)~\cite{muonCosine100} by the COSINE-100 experiment~\cite{cosine100}. 
YEMI is located $\sim$1.5 times deeper than Y2L and, thus, the integrated muon intensities in the units of muons/cm$^{2}$/s for two sites, calculated from Eqs.~(\ref{eq:modgaisser})--(\ref{eq:4}) using both contour maps of Mt. Yemi and Yangyang areas, are found to be as different as 4.6 times. 
We used the measured muon flux at Y2L, by scaling it with 4.6, to derive the muon flux at YEMI with the assumption that their rock properties are equal to each other for two sites. 

The mean muon energy and the vertical muon intensity for YEMI are 236~GeV and 4.0$\times10^{-8}$~muons/cm$^{2}$/s/sr, which are consistent with the measured or simulated values reported in Ref.~\cite{Kudryavtsev:2003}. 

\subsubsection{Brute-force muon simulation}  
\label{subsubsec:3.1.3}
To simulate all the primary and secondary particles induced when the muons interact with rock, the shielding materials, and the detector components, 
we have performed Geant4 Monte Carlo simulations that start with muons, given by the underground differential energy spectrum described in Sects.~\ref{subsubsec:3.1.1} and ~\ref{subsubsec:3.1.2}, by generating them from the outer surface of the rock shell surrounding the cavern, which is called a {\it brute-force} muon simulation.
For the {\it brute-force} muon simulation, we added the rock volume with the thickness that was optimized based on another simulation; we simulated muons with the energy of 236~GeV that is the mean energy of the muon energy spectrum at YEMI by generating it into the rock volume and estimated both of the mean energy and event rate, as a function of the rock thickness, of neutrons and gammas induced by muon interactions with materials in the rock that exited the rock volume. As a result, we added a 3-meter-thick rock shell outside the cavern.
The schematic view of the simulation is shown in Fig.~\ref{bruteforcesim} and one hundred million muons corresponding to a 7-year period were simulated for this study.

\subsection{Results}
\label{subsec:3.2}
We not only analyzed muons but also all the neutrons and gamma rays induced by muon interactions with materials in the rock, in the shield, and in the detector components, as described in Sect.~\ref{subsubsec:3.1.3}. We evaluated the shielding effects with two different shielding configurations using the Muon Veto system to estimate background rates quantitatively. The details of the simulation results are given in Sects.~\ref{subsubsec:3.2.1}, \ref{subsubsec:3.2.2}, and \ref{subsubsec:3.2.3}.  

\begin{figure*}[ht]
\begin{center}
\begin{tabular}{cc}
\includegraphics[width=0.3\textwidth]{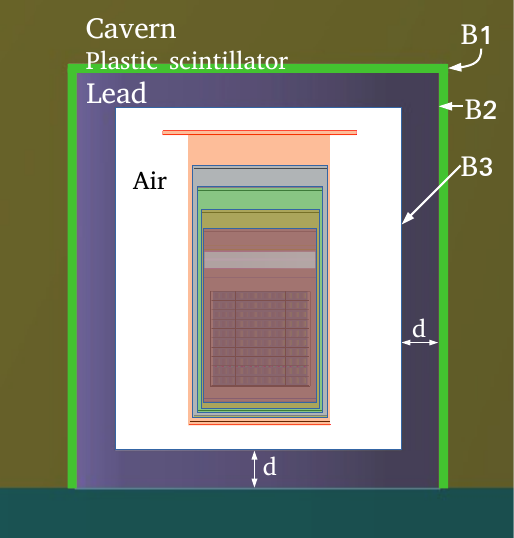} &
\includegraphics[width=0.3\textwidth]{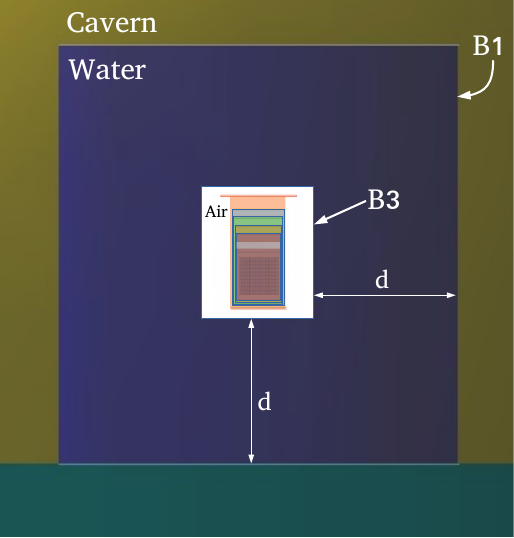} \\
(a) & (b)  \\
\end{tabular}
\caption{
Shielding configurations: (a) lead shield ($d=30 cm$) (b) water shield ($d=3 m$).
}
\label{shieldings}
\end{center}
\end{figure*}

\begin{figure*}[ht]
\begin{center}
\begin{tabular}{cc}
\includegraphics[width=0.5\textwidth]{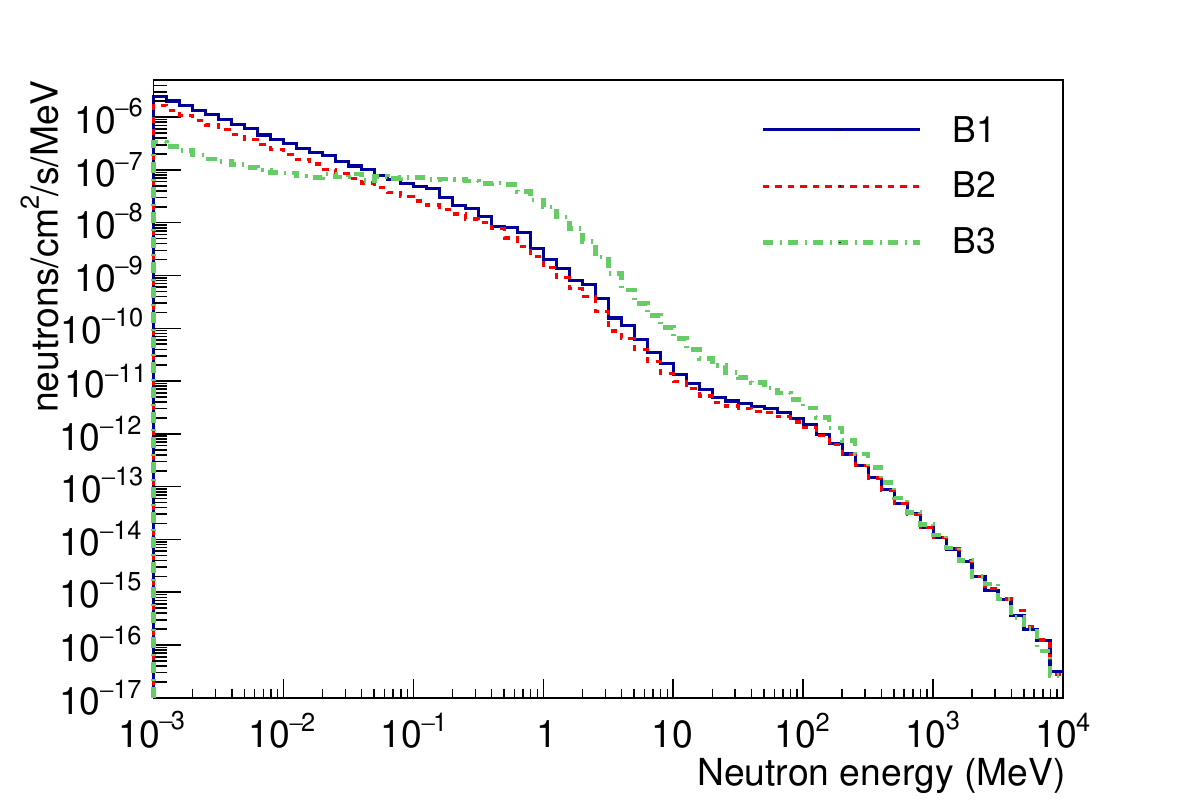} &
\includegraphics[width=0.5\textwidth]{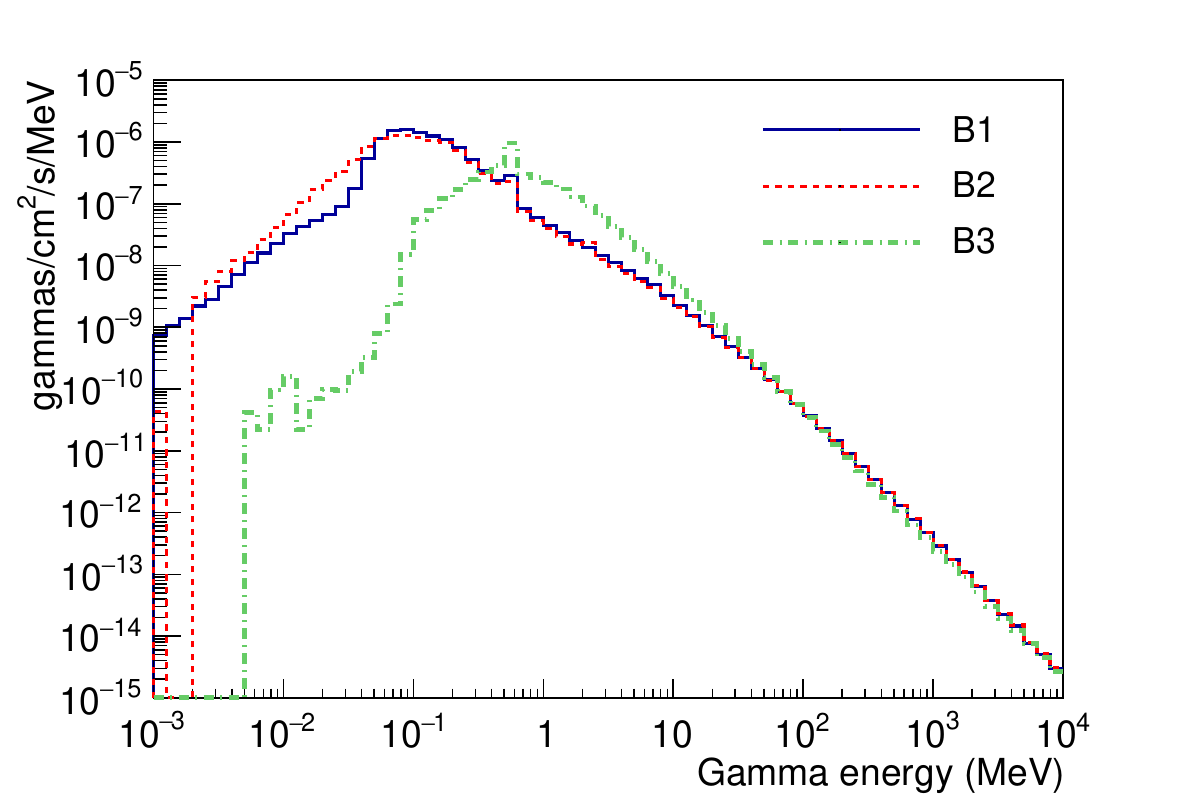} \\
(a) & (b)  \\
\includegraphics[width=0.5\textwidth]{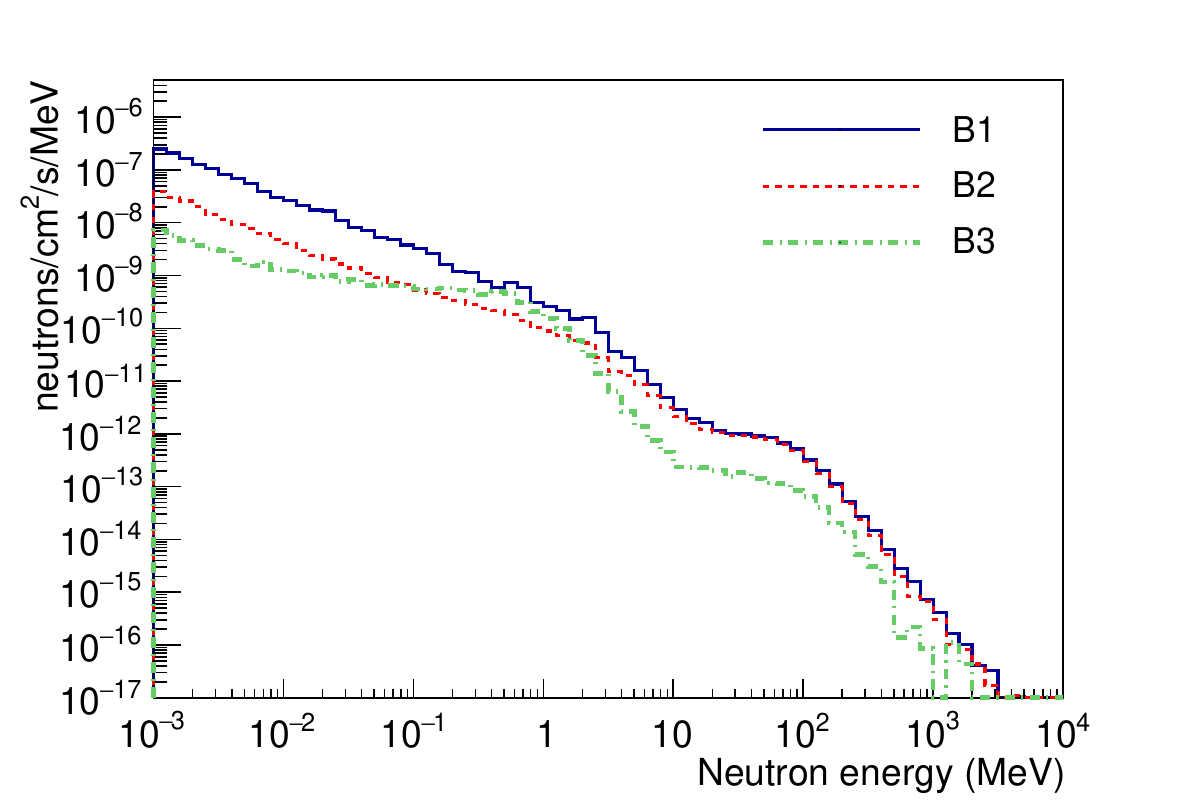} &
\includegraphics[width=0.5\textwidth]{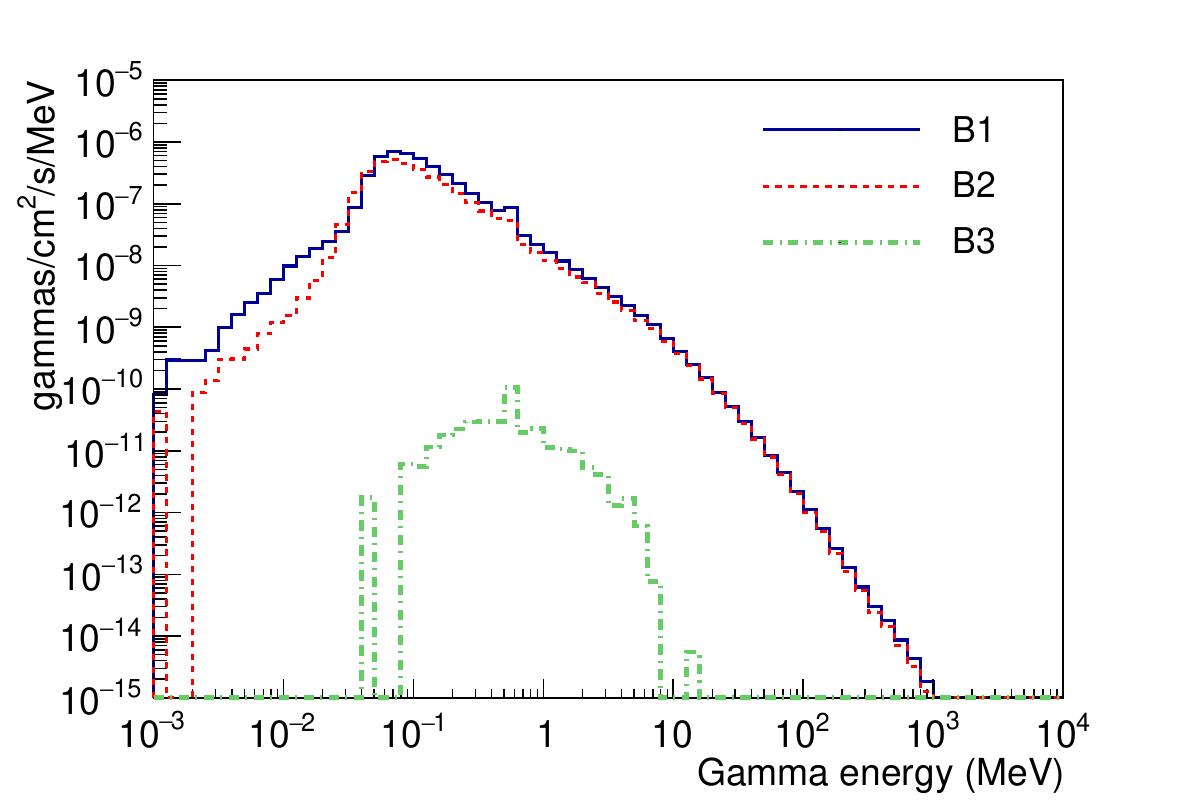} \\
(c) & (d)  \\
\end{tabular}
\caption{
Neutron ((a),(c)) and gamma rays ((b),(d)) energy spectra at the boundaries B1 (black), B2 (red), and B3 (green) with a 30-cm-thick lead shield in configuration 1. (a) and (b) represent the neutron and gamma energy spectra before vetoing muon-tagged events, respectively, and (c) and (d) represent the neutron and gamma energy spectra after vetoing muon-tagged events, respectively. 
}
\label{energyatleadsurface}
\end{center}
\end{figure*}

\subsubsection{Muon-induced neutrons and gamma rays at the boundary between rock and cavern}
\label{subsubsec:3.2.1}
Fig.~\ref{energyatrocksurface}(a) and (b) show the neutron and gamma energy spectra at the rock surface, respectively, produced by the simulation of the muon propagation and its interaction with materials in the rock, described in Sect.~\ref{subsubsec:3.1.3}.
We also included neutrons backscattered from the rock surface, as well as from the shielding materials surrounding the detector that reenter the cavern, represented by the thick solid line of Fig~\ref{energyatrocksurface}(a). According to the simulation results, the muon-induced neutron flux at YEMI is 2.4$\times$10$^{-9}$ n/cm$^2$/s, for energies above 1 MeV. 
The muon-induced neutron flux has a dependence on the rock composition and the depth of the underground laboratory and, thus, it is compared with other similar results. The vertical depth of YEMI (1005~m) is similar to that of Boulby (1070~m) and Boulby's rock composition ($\textless$Z$\textgreater$~=~11.7, $\textless$A$\textgreater$~=~23.6) is similar to the standard rock ($\textless$Z$\textgreater$~=~11, $\textless$A$\textgreater$~=~22)
, as reported in Refs~\cite{Robinson:2003}. 
It is in good agreement with the results of 1.34$\times$10$^{-9}$ n/cm$^{2}$/s (Mei and Hime's prediction) and 8.7$\times$10$^{-10}$ n/cm$^2$/s (FLUKA predciton) at Boulby, values reported in Refs.~\cite{Mei:2005gm,Araujo:2004rv}.

There are features found in the muon-induced gamma-ray spectrum of Fig.~\ref{energyatrocksurface}(b): the peak and the absence of gammas for energies below $\sim$100~keV. It occurred because photoelectric absorption that transfers the energy from a gamma ray to an atomic electron of materials in the rock is more common when the energy of gamma ray is of the same order of magnitude as the binding energy of the atomic electron that is relatively low energy and it interacts with high atomic number materials. Gamma attenuation coefficient for rocks as a function of gamma energy shows strong energy dependence in the range 2--20~keV, dominated by photoelectric absorption, as reported in Ref.~\cite{gammaAtt:2019}.

\subsubsection{Shielding configurations and muon veto system} 
\label{subsubsec:3.2.2}
There are three main sources of backgrounds, found from the brute-force muon simulation, in the cavern: muons, neutrons, and gamma rays. The neutrons and gamma rays are produced by the muons traversing through the rock, as described in the previous Sections. In order to reject the muons and muon-induced backgrounds from the rock surface, we consider the shield layer around the detector. However, neutrons and gamma rays can also be produced by muon interactions with the shielding materials, as well as with the detector components. Thus, we install a Muon Veto system in addition to the shielding materials. 

In this study, we considered two different shielding configurations as shown in Fig.~\ref{shieldings}: a 30-cm-thick lead shield ((a), configuration 1) and a 3-meter-thick water shield ((b), configuration 2). In the simulation, we installed 5-cm-thick plastic scintillator outside the lead shield as a Muon Veto system for configuration 1. For configuration 2, we can measure the energy deposited in the water by using photomultiplier Tubes (PMTs) installed in the water tank. Using the veto system, we can tag muon events that deposit at least as much energy as a minimum energy threshold, and reject all the backgrounds induced by the muon, as muon-tagged events.
We applied the energy threshold of 7.5~MeV to the scintillating veto system in shielding configuration 1 and 500~MeV in configuration 2. In water shielding configuration the energy threshold was changed in terms of the shield thickness. 500~MeV is corresponding to the energy threshold of 3-meter-thick water shield.

We tested the shielding effects by comparing the neutron and gamma energy spectra, produced by muon interaction with rock, at the following boundaries: between the cavern and the plastic scintillator (B1), between the plastic scintillator and the lead (B2), and between the lead/water shields and the air (B3).  
Fig.~\ref{energyatleadsurface} shows the neutron and gamma energy spectra at the boundaries B1, B2, and B3 with configuration 1. The black, red, and green colors represent the neutron/gamma energy spectra at B1, B2, and B3, respectively.  In the simulation, we included a steel skeleton that supports the veto system and the shield layers, affecting the spectrum at B1.

From the neutron spectrum at B3 of Fig.~\ref{energyatleadsurface}(a), we found that the neutrons are built up in the lead shield, similar to results given in Refs.~\cite{Araujo:2004rv, Kudryavtsev:2008, neutronyieldinlead:2013}. There is a small reduction in the neutron backgrounds at B1 and at B2 of Fig.~\ref{energyatleadsurface}(a) by the plastic scintillator layer. That occurred because the neutrons produced by the muon interaction in the scintillator were included. To exclude all the muon-induced backgrounds, we vetoed the muon-tagged events and the resulting neutron/gamma backgrounds at several boundaries are shown in Fig.~\ref{energyatleadsurface}(c) and (d). High-energy neutrons and gamma rays that deposit as much energy as the energy threshold of the veto system, are also rejected as muon-tagged events. Fig.~\ref{energyatleadsurface}(c) represents the neutrons only from the rock or induced by neutrons in the shielding materials. As shown in the two neutron spectra at B1 and at B2 of Fig.~\ref{energyatleadsurface}(c), it is found to be effective to shield neutrons for energies below $\sim$1 MeV with even 5-cm-thick plastic scintillator~\cite{neutronmoderator}. 

\begin{figure}[t]
\begin{center}
\includegraphics[width=0.5\textwidth]{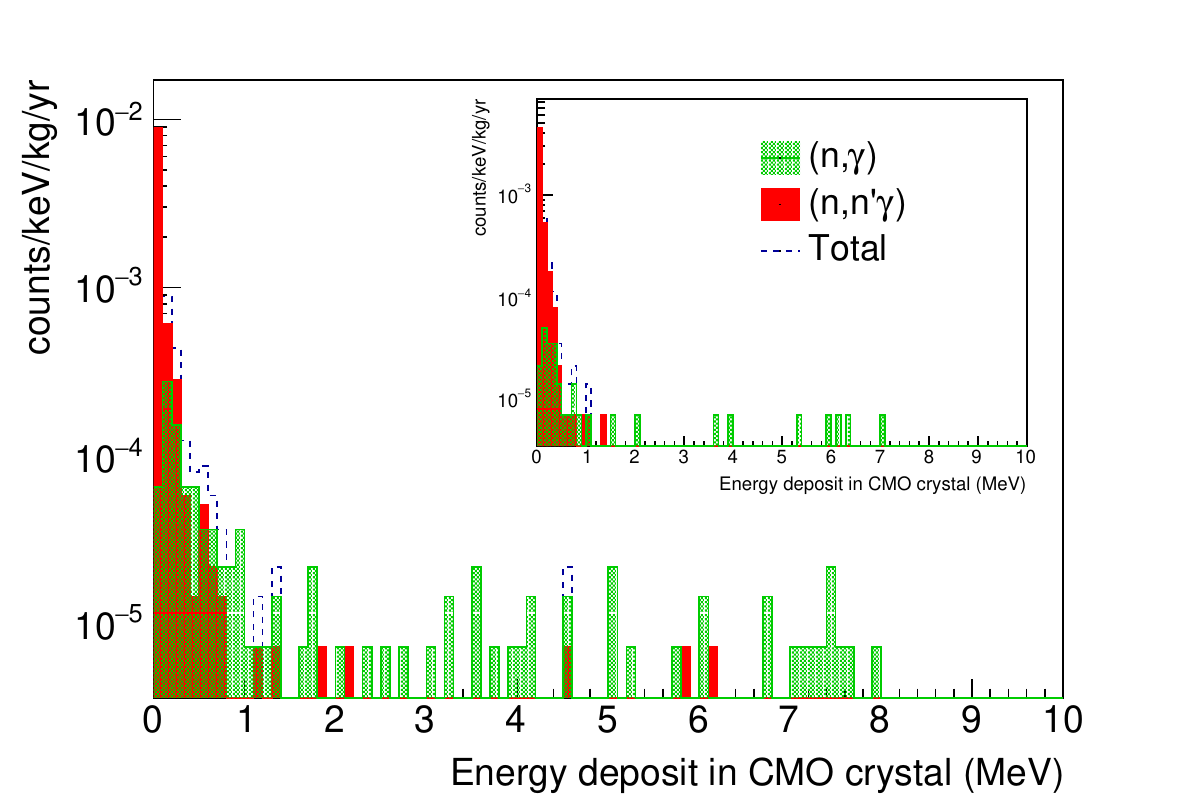} 
\caption{
Energy spectra deposited in a single CMO crystal in configuration 1. It represents the background contributions by (n,~n'$\gamma$) in red and by  (n,~$\gamma$) in green. Single-hit energy spectrum with 50-cm-thick polyethylene shielding installed outside the plastic scintillator in the cavern is shown in the inset. 
}
\label{edepcmo}
\end{center}
\end{figure}

\begin{figure*}[!b]
\begin{center}
\begin{tabular}{cc}
\includegraphics[width=0.5\textwidth]{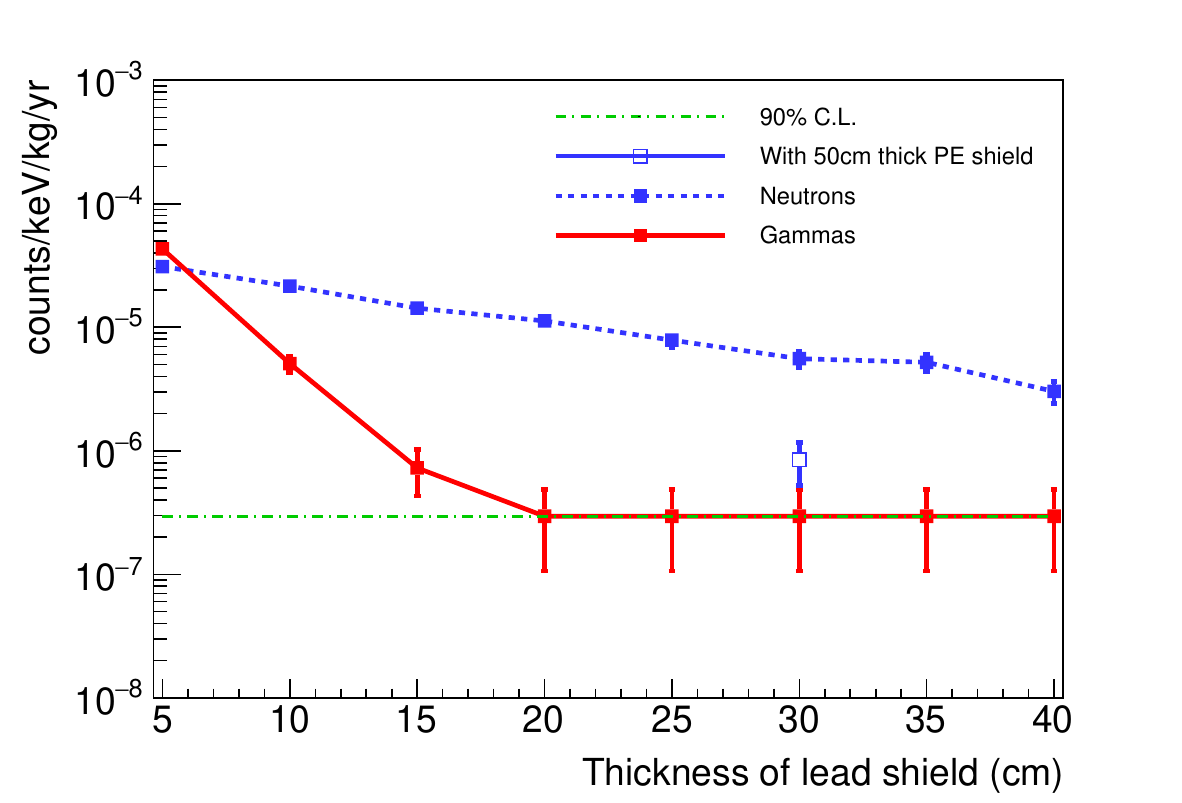} &
\includegraphics[width=0.5\textwidth]{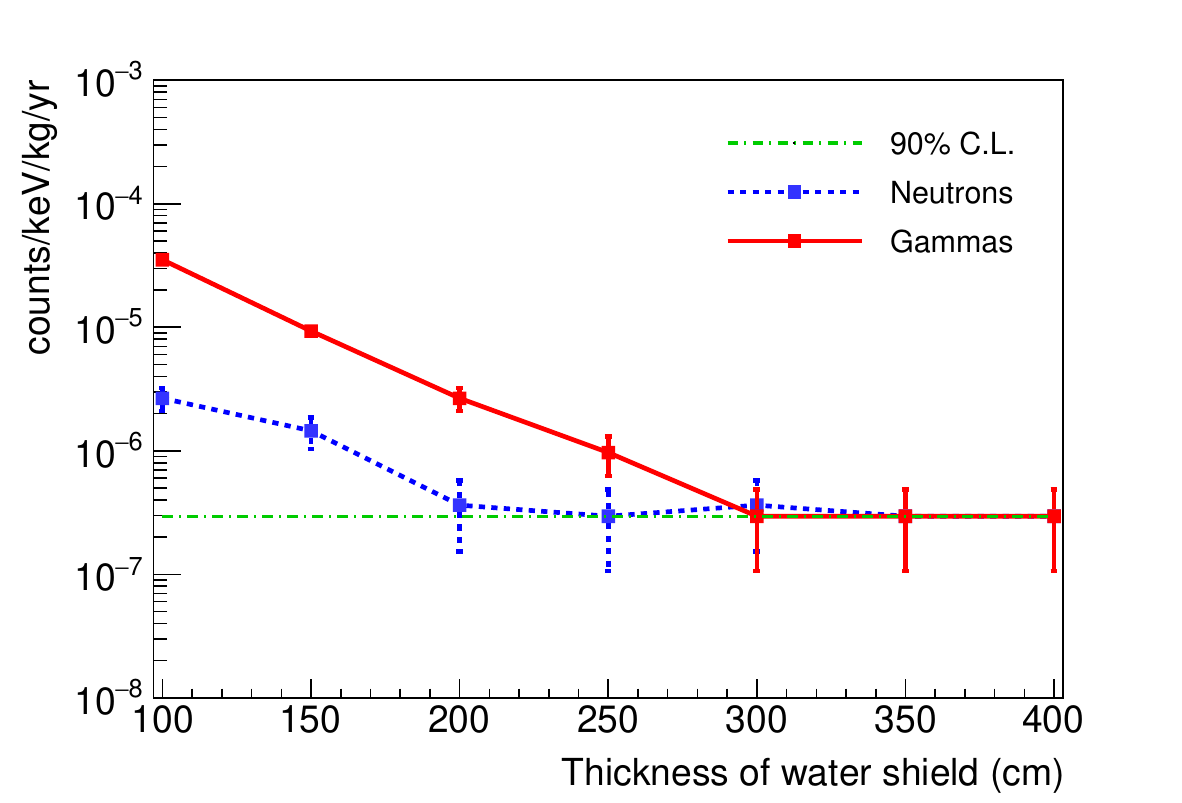} \\
(a) & (b)  \\
\end{tabular}
\caption{
Background rates as a function of the thickness of the shielding materials in (a) configuration 1 and (b) configuration 2.
}
\label{bkgrateleadwater}
\end{center}
\end{figure*}

According to the simulation results of gamma rays in Fig.~\ref{energyatleadsurface}(b), most of the low-energy gamma rays are blocked by the lead shield. However, there is a little reduction for energies above $\sim$10~MeV, even with a 30-cm-thick lead shield. This occurs because a large amount of gamma rays with a few hundreds of keV, to a few hundreds of MeV energy, is formed at B3 (green dashed line). These rays are produced in the lead shield due to the \textit{bremsstrahlung} process, which causes muon-induced electromagnetic showers, tagged by the Muon Veto system, and accordingly rejected as muon-tagged events. The green dashed line of Fig.~\ref{energyatleadsurface}(d) represents gamma energy spectrum at B3 after vetoing muon-tagged events. As a result, the gamma background inside the lead shield, originated at the rock or induced by neutrons in the shielding materials, is found to be negligible in configuration 1.

\subsubsection{Neutron-induced background} 
\label{subsubsec:3.2.3}

In this section, we provided a quantitative understanding of backgrounds induced by neutrons that are induced by muons. 
Fig.~\ref{edepcmo} shows the energy spectrum deposited in a single CMO crystal in configuration 1, called as single-hit events, when considering the neutron flux at B3, shown in the green dotted line of Fig.~\ref{energyatleadsurface}(c), after vetoing the muon-tagged events. We examined how neutrons inside the lead shield make single-hit background events in a crystal. It was found that it occurred via two dominant processes: neutron inelastic scattering (n,~n$^\prime\gamma$) and neutron capture (n,~$\gamma$). We showed their contributions in two different colors in Fig.~\ref{edepcmo}: (n,~n$^\prime\gamma$) in red and (n,~$\gamma$) in green. 

The $\gamma$-rays generated from thermal neutron capture by the (n,~$\gamma$) process in the stainless-steel and copper shields composing the cryostat and the copper material around crystal detectors, contribute mainly to the high-energy backgrounds for energies above $\sim$1~MeV. This includes the 3.034~MeV region of interest (ROI). The background events resulting from (n,~n$^\prime\gamma$) process give only a little contribution for the high energies, contributing mainly to the low-energy region, below $\sim$1~MeV. 

Thus, we tested the shielding effect of neutrons that lead to gamma ray backgrounds by the neutron capture process, using a 50-cm-thick polyethylene (PE) shielding layer installed outside the plastic scintillator in the cavern, and its result is shown in the inset of Fig.~\ref{edepcmo}. We estimated the background rates of single-hit events with and without the PE shield and it resulted in 9.7$\times10^{-7}$~counts/(keV$\cdot$kg$\cdot$yr) and 4.8$\times10^{-6}$~counts/(keV$\cdot$kg$\cdot$yr) in the (2--8)~MeV energy region, respectively. Accordingly, we found that the background is reduced to a level $\sim$5 times lower by adding the 50-cm-thick PE shield in configuration 1. 

\subsubsection{Single-hit background rates}
\label{subsubsec:3.3.3}
In order to find the thickness of the shielding material that meet the background requirement for the AMoRE-II experiment, $\textless10^{-5}$~counts/(keV$\cdot$kg$\cdot$yr), we estimated the single-hit event rate in the (2--8)~MeV energy region with several shielding thicknesses in both configurations, 1 and 2: 5, 10, 15, 20, 25, 30, 35, and 40~cm for the lead shield and 100, 150, 200, 250, 300, 350, and 400~cm for the water shield.
To understand the shielding effect of neutrons and gamma rays quantitatively, we tested the single-hit background rate for both neutrons and gamma rays. 
The resultant single-hit background rates as a function of the shield thickness for both configurations are shown in Fig.~\ref{bkgrateleadwater}(a) and (b); if we found no event in the 2-8 MeV energy region from the simulation we estimated an upper limit at the 90\% C.L.~\cite{feldman}. In Fig.~\ref{bkgrateleadwater}(a), it is found that even a 10-cm-thick lead shield in configuration 1 is effective to shield muon-induced gamma rays, reducing them to the level of $10^{-6}$~counts/(keV$\cdot$kg$\cdot$yr) (red solid line). It also shows the shielding effect of neutrons with respect to several thicknesses of the lead shield (blue dotted line). The background rate was reduced to the aimed level by adding 50-cm-thick PE shield to the 30-cm-thick lead shield in configuration 1 (blue empty marker). Fig.~\ref{bkgrateleadwater}(b) shows the shielding effect of neutrons and gamma rays by the water shield in configuration 2. It is found that even a 100-cm-thick water shield is more effective to shield neutrons. It also shows that the aimed background level can be achieved with 200-cm-thick water shield in configuration 2. In addition, we tested the single-hit rate of muon-tagged events in the (2-4) MeV energy region. The rate was found to be $\sim$10$^{-6}$~counts/(keV$\cdot$kg$\cdot$yr) for both shielding configurations with a 30-cm-thick lead and 3-m-thick water shield when the muon tagging efficiency is 99.9\% 
that meets the background requirement for the AMoRE-II experiment ($\textless10^{-5}$~counts/(keV$\cdot$kg$\cdot$yr)).

\section{Neutron background}
\label{sec:4}
\subsection{Neutron flux at underground laboratory}
\label{subsec:4.1}
The neutron background at YEMI underground laboratory is mainly composed of two types of neutrons. The first contribution is neutrons from the experimental environment, such as neutrons from ($\alpha$,~n) natural radioactivity reactions, and  neutrons from spontaneous fission mainly of U atoms from the underground environment. The second is neutrons induced by muons. The dominant type of neutrons is the environment neutrons because its level is two or three orders of magnitude higher than the muon-induced neutrons~\cite{Carston:2004}. Therefore, we need to test the effect of backgrounds not only from muon-induced neutrons but also from underground environment neutrons, although we evaluated quantitatively the background level of muon-induced neutrons by configuring shielding layers with the Muon Veto system in Sect.~\ref{sec:3}.  

\subsection{Shielding effects and results}
\label{subsec:4.2}
\begin{figure}[t]
\begin{center}
\includegraphics[width=0.5\textwidth]{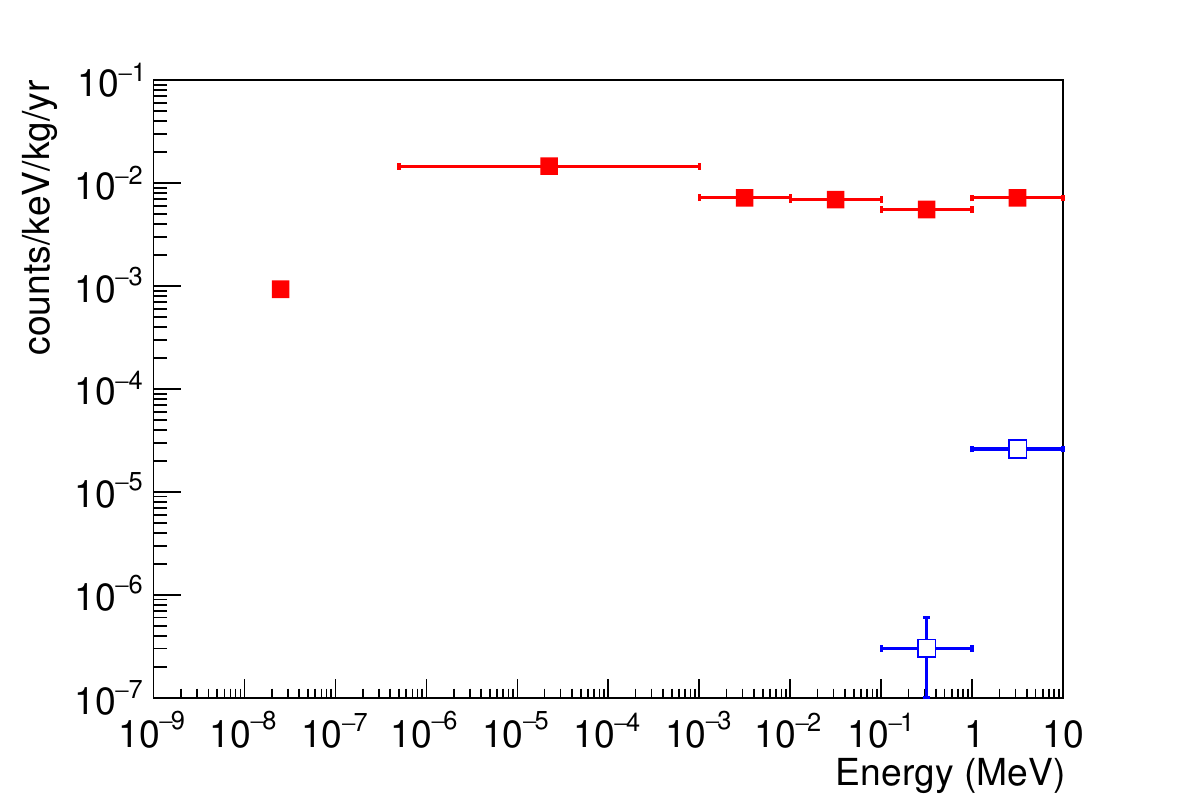} 
\caption{
Neutron background rates with and without 50-cm-thick PE shield in configuration 1. 
}
\label{neutronbkgrate}
\end{center}
\end{figure}

We simulated neutrons by generating them from the rock surface with flat energy spectra over six different energy ranges:  

\begin{enumerate}[(1)]
	\item 0.025~eV 
	\item 0.5~eV -- 1~keV 	
	\item 1~keV -- 10~keV
	\item 10~keV -- 100~keV
	\item 100~keV -- 1~MeV
	\item 1~MeV -- 10~MeV
\end{enumerate}

We used the neutron flux measured by the Bonner sphere spectrometer system at Y2L underground laboratory (Y2L)~\cite{neutronY2L} as a reference value, integrated in a large energy bin and measured in the six different energy binning itemized above. 
However, it needs a more detailed spectrum for some purposes due to the large energy bin width that results in a large uncertainty in the event rate.

The total flux of neutrons at Y2L was measured twice by the Bonner spheres in 2012~\cite{neutronY2L} and by $^{3}$He gas detector in 2018. The results of 6.6$\times$10$^{-5}$~neutrons/cm$^{2}$/s and 4.3$\times$10$^{-5}$~neutrons/cm$^{2}$/s, were found for energies below 10 MeV, respectively. They are consistent with each other, but they are about 10$\sim$17 times higher than that found by Gran Sasso (3.78$\times$10$^{-6}$~neutrons/cm$^{2}$/s~\cite{neutronGranSasso}). 
We assumed that the neutron flux at YEMI is similar to that of Y2L 
where the neutron backgrounds for energies below 10~MeV are dominated by neutrons from the ($\alpha$,~n) process.

The simulation results with configuration 1 for the single-hit event rate in the (2--8) MeV energy region at the six energy binnings are shown in red in Fig~\ref{neutronbkgrate}. It shows the single-hit background rates in the level of 10$^{-3}$~counts/(keV$\cdot$kg$\cdot$yr) for the overall energy range. 

We tested the shielding effects of neutrons using a 50-cm-thick PE shielding layer in the cavern and estimated the background rate for single-hit events in the (2--8) MeV energy region, represented in blue in Fig~\ref{neutronbkgrate}. The background rates of the first four energy binnings (up to 100~keV) are reduced to the level of $\textless10^{-6}$~counts/(keV$\cdot$kg$\cdot$yr) that is an upper limit at the 90\% C.L. with zero-entry; if we found no event in the 2-8 MeV energy region from the simulation we estimated an upper limit at the 90\%~C.L.~\cite{feldman}.  The background rate of the last energy binning of the 1--10 MeV energy interval is found to be about 10$^{-5}$~counts/(keV$\cdot$kg$\cdot$yr) with the PE. 
We found that the background level can be reduced by $\sim$20\% with only an additional 4-mm-thick silicone rubber sheet with 24\% concentrations of boron carbide inside the PE shield from the simulation. 

\section{Conclusion} 
\label{sec:conclusion}
In this study, we have simulated both the cosmic-ray muons and the underground environment neutrons, as well as all the secondary particles. 
Because these backgrounds depend strongly on an experimental design, as reported in Refs.~\cite{gerda:2007,cuore:2010}, 
we quantitatively tested the effects of the neutron and muon-induced backgrounds by configuring different shielding layers with the active Veto system. We studied the shielding effects with various thicknesses of lead in configuration 1 and found that it is good enough to shield muon-induced gamma rays with even a 10-cm-thick lead layer for the AMoRE-II experiment. However, there should be other gammas from the radioisotopes in the rock, which should be considered with a lead layer as thick as 30~cm. 
In addition, an additional 50-cm-thick PE is needed to effectively shield neutrons. 
The water shield in configuration 2 is found to be more effective than lead for shielding neutrons. Therefore, it is also possible to put a 100-cm-thick water shielding layer in configuration 1, instead of the PE shield, to meet the zero-background requirement for the AMoRE-II project.

\section*{Acknowledgments}
This research was funded by the Institute for Basic Science (Korea) under project code IBS-R016-D1 and was supported by the National Research Foundation of Korea(NRF No-2018R1A6A1A06024970) funded by the Ministry of Education, Korea.




\end{document}